# Effects of the free evolution in the Arthurs-Kelly model of simultaneous measurement and in the retrodictive predictions of the Heisenberg uncertainty relations.

J. A. Mendoza-Fierro[1], L. M. Arévalo Aguilar [2], and V. M. Velázquez Aguilar[3]

[1,2]Facultad de Ciencias Físico Matemáticas, Benemérita Universidad Autónoma de Puebla, 18 Sur y Avenida San Claudio, Col. San Manuel, Puebla, Pue., 72520, México

[3] Facultad de Ciencias, Universidad Nacional Autónoma de México, Ciudad Universitaria, D.F. 04510, México

**Abstract**

The simultaneous measurement approach of Arthurs and Kelly has been a significant tool for the better understanding of the measurement process in quantum mechanics. This model considers a strong interaction Hamiltonian by discarding the free evolution part. In this work, we study the effect of the full dynamics –taking into account the free Hamiltonian– on the optimal limits of retrodictive and predictive accuracy of the simultaneous measurement process of position and momentum observables. To do that, we consider a minimum uncertainty Gaussian state as the system under inspection, which allows to carry out an optimal simultaneous measurement. We show that the inclusion of the free Hamiltonian induces a spreading on the probability density of the measurement setting, which increases the value of the product of the variances of the so-called retrodictive and predictive error operators, this is equivalent to a reduction in the accuracy of the measurement.

## 1 Introduction

Measurements in physics constitute the bridge between the theory and its predictions. In this context, the limits of the measurement precision in quantum mechanics are given by the Heisenberg uncertainty relations and the measurement features of quantum mechanics. In classical mechanics, the act of measuring encompass the comparison of a property of a physical object with another acting as a meter; thus, the information obtained from it constitutes an accurate result of the measurement. However, when extending this process to quantum mechanics, the resulting theoretical models, experimentally confirmed, point to different processes with different consequences; for example, in an ideal measurement an entangling system-meter interaction which command the process is required [1].

On the other hand, it is well known that in the quantum world, incompatible observables require their own measurement arrangement which are, generally speaking, incompatible. Some consequences of this situation are that in quantum mechanics the notion of exact accuracy in the measurement process is more restrictive, and restraints on the simultaneous measurement of noncommuting observables arises.

The first person to challenge the classical deterministic conceptions was Heisenberg [2], who heuristically proposes an uncertainty relation $\delta x \delta p \approx \hbar$. However, the widely celebrated uncertainty relation that appears in almost all textbooks of quantum mechanics, i.e.

$$\delta_q \delta_p \geq \frac{\hbar}{2}, \qquad (1.1)$$

was mathematically derived by Kennard [3] who gave a precise meaning to $\delta_q$ and $\delta_p$ as standard deviations.

Today, it is a reached consensus that the inequality given by Eq. ((1.1)) is not related with the simultaneous measurement of complementary observables, rather, it establishes a restriction on quantum state preparation [4–9]. On the other hand, the first quantum-mechanically description where a measuring device interacts with a (pure) system to simultaneous measure its position and momentum observables was proposed by Arthurs and Kelly [10]. They conceived their model as a generalisation of the Von Neumann measurement process [11], where faithful tracking of one single observable is achieved [12]. Notably, they raised the fact that a joint measurement process inevitably entails an induced noise that increases the lower bound of Kennard's uncertainty relation by a factor of $\hbar/2$, that is,

$$\delta_q \delta_p \geq \hbar, \tag{1.2}$$

where now the standard deviations characterize the widths of the probability distributions of the pointer readings of a measuring device. Notably, the Arthurs-Kelly model has led to what is known as pointer-based simultaneous measurements of conjugate observables [13]; besides, it has been considered in interestingly scopes as open systems [13] and decoherence process [14], weak measurements [15], timing in the measurement process [6], entanglement generation [1,16], entanglement swapping, remote tomography and noiseless quantum tracking of conjugate observables [12]. Additionally, there are many experimental accomplishment of simultaneous measurement, see for example references [17–19]

Furthermore, in the Arthurs-Kelly model [20–22] it is assumed that only the interaction Hamiltonian rules the dynamics of the measurement process, neglecting the contribution of the free energies. Hence, this approach focus on the regime where the measuring instruments and the system under investigation are strongly coupled to each other, which is an ideal situation. Besides, using this approach, in previous works was pointed out a necessary distinction between the predictive and retrodictive aspects of a joint measurement [22–25]; building on these ideas, in references [8,26,27] were proposed two kinds of error observables to capture the full content of the experimental accuracy concept in the Arthurs Kelly measurement process. Recent advances in the retrodictive prediction of two incompatible observables, with a precision below the Standard Quantum Limit, was given in references [28,29] using the *past quantum state theory* [30].

In recent years the Arthurs-Kelly model (AKM) of simultaneous measurement, i.e. using the strong coupling regime where the free evolution Hamiltonian is disregarded, has been still used to analyse different characteristics of experimental and theoretical arrangements of simultaneous measurement. For example, Hacohen-Gourgy et al. [31] has carried out a simultaneous measurement of non-commuting observables applying single quadrature measurements using an effective Hamiltonian that resembles the Arthurs-Kelly model, i.e. without the free evolution components. Additionally, Jian and Watanabe [32] also uses the AKM by generalising it for arbitrary operators to obtain a master equation for continuous measurement and for state preparation, notably these authors conclude that the generalised AKM is valid only for weak measurement, i.e. for small values of the coupling constant. Also, remote tomographic reconstruction and teleportations has been proposed by Roy et al. [33] using the AKM. The influence of entangled measuring apparatuses on the AKM model is studied in references [34,35], for an alternative approach about how entanglement resources could improve the simultaneous measurement see reference [29].

Additionally, the AKM model of simultaneous measurement has been extended to other scenarios. For example, it was used to investigate a cloning machine optimised to simultaneous measure non-commuting observable [36]; similarly the uses of quantum cloning to achieve simultaneous measurement was studied in [37]. It was also proposed by D'Ariano et al. to simultaneously measure the direction of the spin [38], see also Martens and de Muynick [39]. The AKM has been also used to model macroscopic measurement [40]. In the experimental realm of the new scenarios, there has been recent interesting realizations of the simultaneous measurement of incompatible observables, in references [41–43] the joint measurability of noncommuting polarizations observables was theoretically analized through the error statistics and the nonclassical correlation between measurement errors was experimentally measured. As a highly important result, it was shown that to enforce restrictions on experimental probabilities to be positive (in joint measurements) defines a bound on non-local correlations [43].

On the other hand, the simultaneous measurement of incompatible observables was experimentally implemented in a pair of polarization components of a single photon avoiding the entanglement with a measurement apparatus [44]; here the authors shows a simultaneous measurement technique that avoids coupling the quantum system to an ancillary system, showing and alternative to the usual Arthurs-Kelly model.

In this work, we study the effects of the full dynamics in the simultaneous measurement of position and momentum observables; this is done by using a minimum uncertainty Gaussian state as the state under inspection, which allows measuring with optimal accuracy. The analysis of the dynamics of the system is made in the Schrödinger picture by using the time evolution operator method [45]. We found that the effects of the free terms of the Hamiltonian is to increase the noise in the measurement, this effect is done by increasing the lower bound in the uncertainty relation. We also show that the uncertainty relation that we found approach the uncertainty relation found by Arthurs and Kelly in the limit of strong coupling.

This contribution uses $\hbar = 1$ and is organized as follows: in Sec. 2, we define the measurement configuration. Besides, we use the time evolution operator method to obtain the state describing the coursing dynamics; by using this result, we compute the variances of the probability distributions of the pointers and system just after the time of the measurement. By using these results, in Sec. 3 we show how the full dynamics affect the retrodictive and predictive aspects of accuracy in an optimal simultaneous measurement. The paper close with the conclusions in Sec. 4.

## 2 The measurement process

This section describes the scheme of simultaneous measurement for the position and momentum observables following the scheme used by Arthurs and Kelly [10]. We consider the stage known as pre-measurement [46], where a correlation between the measuring instrument and the system arises; later on, we obtain the variances of the probability distributions of the pointers readings at the time just after the simultaneous readout.

### 2.1 The measurement configuration

The setup that we will consider is the same that the one used by Arthurs and Kelly except for the addition of the free Hamiltonian [10]; that is to say, there is a measuring device formed by two measuring system coupled to a third single system of which we are interested in, we are interested in measuring two non-commuting observables $\hat{x}_3$ and $\hat{p}_3$ [10] of the latter system. The Hamiltonian for this setup is given by:

$$\hat{H} = \hat{H}_\text{int} + \hat{H}_\text{free}, \tag{2.1}$$

where the interaction Hamiltonian $\hat{H}_\text{int}$ is the original of Arthurs and Kelly

$$\hat{H}_\text{int} = \kappa \left( \hat{x}_3 \hat{p}_1 + \hat{p}_3 \hat{p}_2 \right), \tag{2.2}$$

where $\kappa$ is a positive constant that governs the coupling strength between the pointer and its linked variable; the measuring systems are represented by $\hat{p}_1$ and $\hat{p}_2$. Further, we take into account the free dynamics which has not been taken into account by Arthurs and Kelly, by considering

$$\hat{H}_\text{free} = \frac{\hat{p}_1^2}{2m_1} + \frac{\hat{p}_2^2}{2m_2} + \frac{\hat{p}_3^2}{2m_3}. \tag{2.3}$$

The initial state of each system is given by the following wave function:

$$\langle x_i | \phi_i \rangle = \phi_i(x_i) = \frac{\sqrt{S_i}}{\pi^{\frac{1}{4}}} \exp\left[ -\left( x_i S_i \right)^2 / 2 \right], \tag{2.4}$$

where in the remainder of this work the subscript $i = 1, 2, 3$ will label the variables of the pointer 1, the pointer 2, and the system, respectively. The variances of the probability distributions associated with the wave function, Eq. (2.4), are given by $\sigma_i^2 = (S_i)^{-2}/2$, with $S_1 = (2/b)^{\frac{1}{2}}$, $S_2 = (2b)^{\frac{1}{2}}$ and $S_3 = 1/(2^{\frac{1}{2}} \delta_q)$, therefore the resolution of the pointers can be balanced through the manipulation of the $b$ parameter; hence, it is possible to tune up the measurement accuracy for any of the two conjugate observables [6,21]; moreover, it is easy to verify that the position probability distribution of the Gaussian system has a variance of $\delta_q^2$ and that of the conjugate distribution of $\delta_p^2 = 1/4\delta_q^2$.

The Hilbert space of the whole system is represented by the tensor product of the individual Hilbert spaces $\mathcal{H} = \otimes_{i=1}^{3} \mathcal{H}_i$, where each one has an infinitely dimensional structure; then, the initial quantum state is

$$\psi(x_1, x_2, x_3, t=0) = \phi_1(x_1)\phi_2(x_2)\phi_3(x_3). \tag{2.5}$$

This wave function characterises the complete system before the measurement process and is a pure and separable state.

Interestingly, the wave functions given by Eq. (2.4) represent a particular class of minimum uncertainty states known as squeezed vacuum states with squeezing factor $S_i$ [47]. Within these considerations, a full Gaussian minimum uncertainty configuration characterises the whole measurement setting. This representation is suitable in the first instance given the versatility of Gaussian states and its proliferation in practically any field of quantum mechanics. Moreover, these are states which are easily of preparation and control through current technology [48–50], therefore, allowing easy implementation in quantum communication protocols [49,51–53] and they allow reaching the optimal limits of accuracy in a simultaneous measurement of position and momentum observables

## 2.2 The dynamics

At determined time, the pointers of the measurement devices interact with the system of interest to extract the information of the non-commuting observables. Although the laws of quantum mechanics prevent to perform simultaneous measurement of the position and momentum observables with arbitrary accuracy, it is still possible to make an estimate of them by inducing a correlation with the observables of the pointers. Hence, the interaction Hamiltonian is usually represented by

$$\hat{H}_{\text{int}} = \kappa \left( \hat{x}_3 \hat{p}_1 + \hat{p}_3 \hat{p}_2 \right), \tag{2.6}$$

where $\kappa$ is the coupling strength between the pointers of the measurement device and the non-commuting observables of the system. Additionally, we take into account the free dynamics by considering

$$\hat{H}_{\text{free}} = \frac{\hat{p}_1^2}{2m_1} + \frac{\hat{p}_2^2}{2m_2} + \frac{\hat{p}_3^2}{2m_3}, \tag{2.7}$$

with $m_i$ the mass of each constituent of the whole system. Therefore, the total Hamiltonian is

$$\hat{H} = \hat{H}_{\text{int}} + \hat{H}_{\text{free}}. \tag{2.8}$$

This Hamiltonian, i.e. Eq. (2.8), describes the dynamics of the measurement process; notice that, unlike previous works [10, 20–22], the Eq. (2.8) allow us to consider both situations: the strong and weak couplig. It is worthy of mention that the situation of strong coupling, that arises when the free Hamiltonian is discarded [10, 20–22], cannot be sustained in a physically realistic measurement scenario since a completely infinite coupling is an ideal situation, se Eq. (2.20) below.

We obtain the temporal evolution of the initial state, Eq. (2.5), in the Schrödinger picture through the time evolution operator method [45], that is,

$$\Psi(x_1, x_2, x_3, t) = e^{-i\hat{H}t}\psi(x_1, x_2, x_3, t=0). \tag{2.9}$$

Because the Hamiltonian is time-independent, the unitary operator $e^{-i\hat{H}t}$ can be factorized as (see Appendix A)

$$e^{-i\hat{H}t} = e^{\Delta x_1 \hat{p}_1^2} e^{-\frac{it}{2m_2}\hat{p}_2^2} e^{-\frac{it}{4m_3}\hat{p}_3^2} e^{-\frac{it\kappa}{2}\hat{p}_3\hat{p}_2} e^{-it\kappa \hat{x}_3 \hat{p}_1} e^{-\frac{it}{4m_3}\hat{p}_3^2} e^{-\frac{it\kappa}{2}\hat{p}_3\hat{p}_2}, \tag{2.10}$$

The application of the time evolution operator to the initial wave function is shown in the appendix; here we present the result in compact form as:

$$\Psi(\hat{\chi}, t) = \mathcal{N}(t) \exp\left[-\left\{\varepsilon_1(t)x_1^2 + \varepsilon_2(t)x_2^2 + \varepsilon_3(t)x_3^2 + \varepsilon_4(t)x_1x_2 + \varepsilon_5(t)x_1x_3 + \varepsilon_6(t)x_2x_3\right\}\right], \tag{2.11}$$

where $\mathcal{N}(t)$ and $\varepsilon_j(t)$ are complex time-dependent functions defined in the Appendix B. The wave function given by Eq. (2.11) has associated a three-variable Gaussian probability distribution; besides, it can be verified that it is normalized for all $t$. Furthermore, it describes the dynamics of the pointers plus the system while the measurement is in progress. Since it cannot be expressed as a product of individual functions, it is an example of three-mode Gaussian entanglement [54] between the pointers and system variables [55]. The temporal dynamics of the mean value for any position and momentum function $Q(\hat{\chi}, \hat{\varrho}, t)$ can be obtained through

$$\langle Q(\hat{\chi}, \hat{\varrho}, t) \rangle = \int \Psi^*(\chi, t) \, Q(\hat{\chi}, \hat{\varrho}, t) \, \Psi(\chi, t) \, d\chi, \tag{2.12}$$

with $\hat{\chi}$ ($\hat{\varrho}$) summarising the position (momentum) dependence of the function, hence $d\chi = \prod_{i=1}^{3} dx_i$.

For the remaining of this paper, we take the mean value of the $n$-th moment of some observable at the time in which the measurement is carried out, i.e. as the reciprocal of the coupling constant $t = \tau = 1/\kappa$ [10] [1], this election simplify the definitions and calculations. As usual, the variance is defined as $\sigma_{\hat{A}}^2 = \left\langle \hat{A}^2 \right\rangle - \left\langle \hat{A} \right\rangle^2$. We emphasise that each marginal probability distribution associated with the wave function given by Eq. (2.11) is Gaussian with zero mean value, therefore, their variances coincide with the squared root-mean-square (rms) error [57], which is a reasonable measure of dispersion, it quantifies the degree at which the probability distribution of an observable deviates from another that is pretended to estimate [58,59].

Using the wave function, i.e. Eq. (2.11), and the definition given in Eq. (2.12), we find that the $\kappa$-dependent variances of the position probability distributions of the pointers at the time of the measurement are given by

$$\sigma_{\hat{x}_1}^2(\kappa) = \delta_q^2 + \frac{b}{2} + \eta_1(\kappa) + \eta_2(\kappa), \tag{2.13}$$

$$\sigma_{\hat{x}_2}^2(\kappa) = \delta_p^2 + \frac{1}{2b} + \eta_3(\kappa), \tag{2.14}$$

while the variances of the conjugate distributions of the Gaussian system at the same time are

$$\sigma_{\hat{x}_3}^2(\kappa) = \delta_q^2 + b + \eta_2(\kappa), \tag{2.15}$$

$$\sigma_{\hat{p}_3}^2(\kappa) = \delta_p^2 + \frac{1}{b}, \tag{2.16}$$

where the $\eta_j(\kappa)$-functions represent the contribution given by the free evolution of the sub-systems of the whole system which was not taken into account before in references [10,20–22], they represent one of the main results of this paper and are defined as:

$$\eta_1(\kappa) = \frac{(m_1 - 6m_3)^2}{36 b m_1^2 m_3^2 \kappa^2}, \tag{2.17}$$

$$\eta_2(\kappa) = \frac{1}{16 \delta_q^2 m_3^2 \kappa^2}, \tag{2.18}$$

$$\eta_3(\kappa) = \frac{b}{m_2^2 \kappa^2}. \tag{2.19}$$

It is worth mentioning that they vanish as the coupling $\kappa$ becomes larger, that is,

$$\lim_{\kappa \rightarrow \infty} \eta_j(\kappa) = 0, \tag{2.20}$$

in this limit the strong coupling regime is recovered and the variances of the probability distributions calculated in [10] are recovered also.

---

[1] At this time the simultaneous readout takes place; therefore, projective measurements are done in the pointers, leaving the post-measurement Gaussian state as the normalised projection on the eigen-space of the observed eigenvalues according to the quantum mechanically postulate of a projective measurement [56].

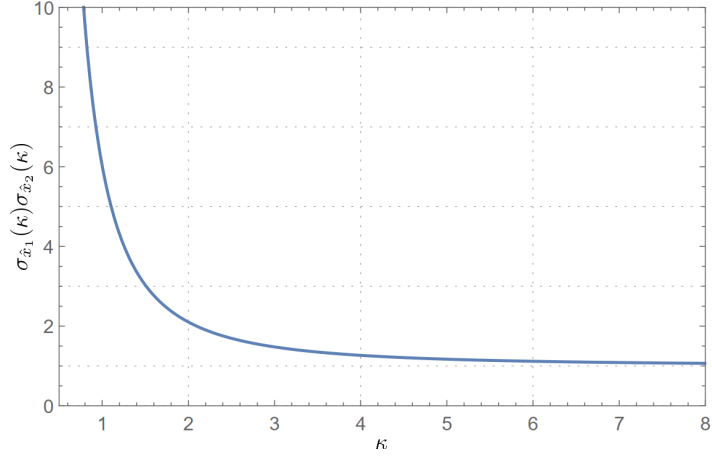

Figure 1: (color online) Behavior of the product $\sigma^2_{\hat{x}_1}(\kappa)\sigma^2_{\hat{x}_2}(\kappa)$ given by Eq. (2.21) versus the coupling constant $\kappa$, for the values $\delta_q = 1$, $m_i = 1$. As the coupling strength $\kappa$ decreases, the product $\sigma^2_{\hat{x}_1}(\kappa)\sigma^2_{\hat{x}_2}(\kappa)$ shift upwards from its minimal value due to the contribution of the $\Delta_1(\kappa)$-function.

In their seminal work, Arthurs and Kelly focused on adjusting the balance parameter $b$ in order to minimise the product $\sigma^2_{\hat{x}_1}\sigma^2_{\hat{x}_2}$; thus, they found that it cannot take a value less than 1 (in units of $\hbar^2$), with the balance parameter of the pointers adjusted at the rate $b = \delta_q \delta_p^{-1}$. For the particular measurement setting here considered, this value is given by $b = 2\delta_q^2$; hence, using it in the product of variances given by Eqs. (2.13) and (2.14) and together with the lower bound of Kennard uncertainty relation Eq. (1.1), the product $\sigma^2_{\hat{x}_1}(\kappa)\sigma^2_{\hat{x}_2}(\kappa)$ in the full dynamical simultaneous measurement process, i.e. when taking into account the free Hamiltonian, is

$$\sigma^2_{\hat{x}_1}(\kappa)\sigma^2_{\hat{x}_2}(\kappa) = \sigma^2_{\hat{x}_1}\sigma^2_{\hat{x}_2}|_{\min} + \Delta_1 = 1 + \Delta_1(\kappa), \tag{2.21}$$

where $\sigma^2_{\hat{x}_1}\sigma^2_{\hat{x}_2}|_{\min} = 1$ and the $\Delta_1(\kappa)$-function is given by

$$\Delta_1(\kappa) = \frac{11m_1^2\left(4\delta_q^4 + \kappa^2 m_2^2\right) - 24m_1\left(4\delta_q^4 + \kappa^2 m_2^2\right)m_3 + 72\left(4\delta_q^4 + \kappa^2\left[16\delta_q^8 m_1^2 + m_2^2\right]\right)m_3^2}{288\left(\delta_q^2 \kappa^2 m_1 m_2 m_3\right)^2}. \tag{2.22}$$

The $\Delta_1(\kappa)$ summarises the contribution of the free evolution to the measurement setting, which will undergo a hyperbolic shift upwards as the coupling strength $\kappa$ becomes smaller; see Fig. 1. However, although the product $\sigma^2_{\hat{x}_1}(\kappa)\sigma^2_{\hat{x}_2}(\kappa)$ gives a quantitative notion of the accuracy in the simultaneous measurement, it only refers to the retrodictive aspect –to be defined in the next section–, besides the variances $\sigma^2_{\hat{x}_1}(\kappa)$ and $\sigma^2_{\hat{x}_2}(\kappa)$ cannot be directly interpretable as experimental errors [8,26].

In the following, we will show how these free energy contribution, i.e. Eqs. (2.17) to (2.19), affects the *accuracy* in the simultaneous measurement, whose meaning we define in the next section through the so-called retrodictive and predictive error operators.

## 3 Free energy contributions to the accuracy of the measurement

To interpret the results of the last section, let us first recall the definitions for the retrodictive and predictive error operators as they were given in [8,26,27]. It is important to note that these definitions can get meaning only when there is a large quantity of measurements on identically prepared systems, such that it forms a probability distribution from which the n-th moment of
6

canonical observables is estimated [21]. Such interpretation is necessary due to the contextuality in quantum mechanics, which asserts that observables do not have a predetermined value before a measurement [60–62].

### 3.1 Error operators and optimal joint measurements

The formalism of quantum mechanics limits the simultaneous measurement of non-compatible observables; nevertheless, this restriction does not exclude the possibility of building inferences about such quantities within certain error margins. Hence, it is implicit that reducing such errors the measurement accuracy is increased. To clarify this conception, assume a measuring apparatus equipped with an initial pointer variable $\hat{x}$ which is coupled through unitary dynamics $\hat{U}$ with the observable $\hat{A}$ which is intended to be measured; it is said that the measurement is carried out with total accuracy if the probability distribution of the pointer variable just after the measurement process coincide with the pre-measurement (or post-measurement, as we will see in short) probability distribution of $\hat{A}$; see for example the definition given by in [63].

Within the context of the simultaneous measurement raised by Arthurs and Kelly, it has been pointed out two aspects of accuracy [8, 26, 27], which quantify the deviation of the probability distributions of the pointer records from those of the canonical pair before as well as after the measurement process. In order to study these aspects in the model given by Eq. (2.8), i.e. taking in to account the free Hamiltonian of the subsystems, in the following we review their conceptual definitions.

The retrodictive error operators are defined, se the work of Appleby [8, 26, 27], as:

$$\hat{\epsilon}_{x_i} \equiv \hat{U}^\dagger \hat{x}_1 \hat{U} - \hat{x}_3, \tag{3.1}$$

$$\hat{\epsilon}_{p_i} \equiv \hat{U}^\dagger \hat{x}_2 \hat{U} - \hat{p}_3, \tag{3.2}$$

where in these and all subsequent definitions, the unitary operator $\hat{U}$ is the one governing the dynamics of the measurement process; hence $\hat{U}^\dagger \hat{A} \hat{U}$ describes the dynamics of the observable $\hat{A}$ in the Heisenberg picture. Through a proper measure of dispersion, the retrodictive error operators give information about how the position probability distributions of the pointers, just after the time of measurement, deviate from the initial probability distributions of the conjugate pair of observables being measured; thus, they have a retrodictive character in the sense that they establish comparison with the information of the observables under inspection before to the measurement process.

The predictive error operators are defined as [8, 26, 27]:

$$\hat{\epsilon}_{x_f} \equiv \hat{U}^\dagger \left( \hat{x}_1 - \hat{x}_3 \right) \hat{U}, \tag{3.3}$$

$$\hat{\epsilon}_{p_f} \equiv \hat{U}^\dagger \left( \hat{x}_2 - \hat{p}_3 \right) \hat{U}, \tag{3.4}$$

which, through an adequate measure of dispersion, they provide information about how the probability distributions of the pointers deviate from those of the observables under inspection, both just after the time of measurement; therefore, have a predictive character in the sense that they pretend to establish a comparison with information of the canonical pair under examination after the measurement process.

A simultaneous measurement process of position and momentum is called retrodictively optimal or with maximal retrodictive accuracy if it is satisfied that [64]

$$\langle \hat{\epsilon}_{x_i} \rangle = \langle \hat{\epsilon}_{p_i} \rangle = 0; \tag{3.5}$$

besides the product of variances associated with the retrodictive error operators is minimum, that is,

$$\left\{ \sigma^2_{\hat{\epsilon}_{\hat{x}_i}} \sigma^2_{\hat{\epsilon}_{\hat{p}_i}} \right\}_{\min} = \frac{1}{4}. \tag{3.6}$$

The condition, Eq. (3.5) means that the mean value of the records of the pointers match with the mean value of the initial (before any interaction with the measuring instrument) observables of the system.

A simultaneous measurement process of position and momentum is called predictively optimal or with maximal predictive accuracy if the product of the variances associated with the retrodictive error operators is minimum [64], that is,

$$\left\{\sigma^2_{\hat{\epsilon}_{\hat{x}_f}} \sigma^2_{\hat{\epsilon}_{\hat{p}_f}}\right\}_{\min} = \frac{1}{4}, \tag{3.7}$$

consequently it is satisfied that

$$\langle \hat{\epsilon}_{x_f} \rangle = \langle \hat{\epsilon}_{p_f} \rangle = 0. \tag{3.8}$$

The condition, Eq. (3.8), means that the pointer's outputs match on average with the mean value of position and momentum observables of the system at the time just after the measurement is carried out.

If the conditions given by Eqs. (3.5) and (3.8) are fulfilled, it can be proved that [8]

$$\left\langle \hat{\epsilon}_A \hat{B} \right\rangle = \left\langle \hat{B} \hat{\epsilon}_A \right\rangle = 0, \quad A \in \{x_i, p_i, x_f, p_f\}, \quad \hat{B} \in \{\hat{x}_3, \hat{p}_3\}, \tag{3.9}$$

where $A$ is the subscript labellings the error operators and $\hat{B}$ represent any of the initial conjugate observables of the Gaussian system to be measured. The last equation will be auxiliary for future calculations.

The measurement instrument considered by Arthurs and Kelly was designed to maximize the retrodictive and predictive aspects of accuracy [8], which become superior within the regime of strong coupling when the system under measurement is a minimum uncertainty Gaussian state. In what follows, we will show that the maximal accuracy on the two aspects of the measurement is not maintained when it is considered the full dynamics of the measurement process –i.e. when the free evolution of the sub-systems is taken in to account– since it introduces an extra noise on the variances of the error operators that depends on the degree of coupling between the measurement instrument and the system to be measure.

## 3.2 The accuracy in the simultaneous measurement

To begin with, using the evolution given by the Eq. (2.11) we can deduce –by using the definitions Eqs. (3.1) and (3.2) and the conditions Eqs. (3.5) and (3.9)– that the variances of the retrodictive error operators are

$$\sigma^2_{\hat{\epsilon}_{x_i}} = \left\langle \left(\hat{U}^\dagger \hat{x}_1 \hat{U}\right)^2 \right\rangle - \langle \hat{x}_3^2 \rangle = \sigma^2_{\hat{x}_1} - \delta_q^2, \tag{3.10}$$

$$\sigma^2_{\hat{\epsilon}_{p_i}} = \left\langle \left(\hat{U}^\dagger \hat{x}_2 \hat{U}\right)^2 \right\rangle - \langle \hat{p}_3{}^2 \rangle = \sigma^2_{\hat{x}_2} - \delta_p^2. \tag{3.11}$$

Considering the definitions Eqs. (3.3) and (3.4), and the condition, Eq. (3.8), the variances of the predictive operators are

$$\sigma^2_{\hat{\epsilon}_{x_f}} = \left\langle \hat{U}^\dagger (\hat{x}_1 - \hat{x}_3)^2 \hat{U} \right\rangle = \sigma^2_{\hat{x}_1} + \sigma^2_{\hat{x}_3} - 2 \left\langle \hat{U}^\dagger \hat{x}_1 \hat{x}_3 \hat{U} \right\rangle, \tag{3.12}$$

$$\sigma^2_{\hat{\epsilon}_{p_f}} = \left\langle \hat{U}^\dagger (\hat{x}_2 - \hat{p}_3)^2 \hat{U} \right\rangle = \sigma^2_{\hat{x}_2} + \sigma^2_{\hat{p}_3} - 2 \left\langle \hat{U}^\dagger \hat{x}_2 \hat{p}_3 \hat{U} \right\rangle, \tag{3.13}$$

where we have use the fact that $[\hat{x}_1, \hat{x}_3] = [\hat{x}_2, \hat{p}_3] = 0$ and $\hat{U}^\dagger \hat{U} = \hat{I}$. The last term in Eqs. (3.12) and (3.13) express the double covariance between the observables that define the predictive error operator; they are obtained in the Schrödinger picture by employing the Eq. (2.12), then

$$2\left\langle \hat{U}^\dagger \hat{x}_1 \hat{x}_3 \hat{U} \right\rangle = 2\delta_q^2 + b + 2\eta_2(\kappa), \tag{3.14}$$

$$2\left\langle \hat{U}^\dagger \hat{x}_2 \hat{p}_3 \hat{U} \right\rangle = \frac{1}{b} + 2\delta_p^2. \tag{3.15}$$

Using Eqs. (2.13) and (2.14) in Eqs. (3.10) and (3.11), the $\kappa$-dependent variances of the retrodictive error operators at the time of the joint measurement are

$$\sigma^2_{\hat{\epsilon}_{x_i}}(\kappa) = \frac{b}{2} + \eta_1(\kappa) + \eta_2(\kappa), \tag{3.16}$$

$$\sigma^2_{\hat{\epsilon}_{p_i}}(\kappa) = \frac{1}{2b} + \eta_3(\kappa). \tag{3.17}$$

Using Eqs. (2.13) to (2.16) and Eqs. (3.14) and (3.15) in Eqs. (3.12) and (3.13), the $\kappa$-dependent variances of the predictive error operators at the same time are

$$\sigma^2_{\hat{\epsilon}_{x_f}}(\kappa) = \frac{b}{2} + \eta_1(\kappa), \tag{3.18}$$

$$\sigma^2_{\hat{\epsilon}_{p_f}}(\kappa) = \frac{1}{2b} + \eta_3(\kappa). \tag{3.19}$$

From the variances of the pointer readings, Eqs. (2.13) and (2.14), it is inferred that the simultaneous measurement would be exact if they were only given by the initial variances of position and momentum probability distributions of the Gaussian state, i.e., $\delta_q^2$ and $\delta_p^2$ respectively; therefore, the extra terms correspond to the noise affecting the retrodictive accuracy of the measurement.

Such noise is quantitatively expressed by the variances of the retrodictive error operators, Eqs. (3.16) and (3.17). In the same way, the noise deviating the probability distributions of the pointer readings from those of the canonical pair just after the time of the measurement is represented by the variances of the predictive error operators, Eqs. (3.18) and (3.19). From the conditions given by Eqs. (3.6) and (3.7) and the variances of the error operators, Eqs. (3.16) to (3.19), it is deduced that the joint measurement would be retrodictive and predictively optimal if the $\eta_j(\kappa)$-functions were zero (which only happens in the strong coupling limit, as is shown by Eq. (2.20)), these terms only appear when the free energy operators in the Hamiltonian of the measurement process is taken into account, they arise as a consequence of the spread of the probability density of the whole Gaussian system.

Nevertheless, it must be noted that even in the optimal situation of accuracy, there exists an unavoidable noise which is proportional to the pre-measurement variances of the probability distributions of the pointers (that is, the $b$-dependent terms appearing in Eqs. (3.16) to (3.19)), which is consequence of the unitary dynamics associated with the quantum measurement process [21,65] and it is the cause for the increment of the lower bound of the Kennard uncertainty relation as was first predicted by Arthurs and Kelly.

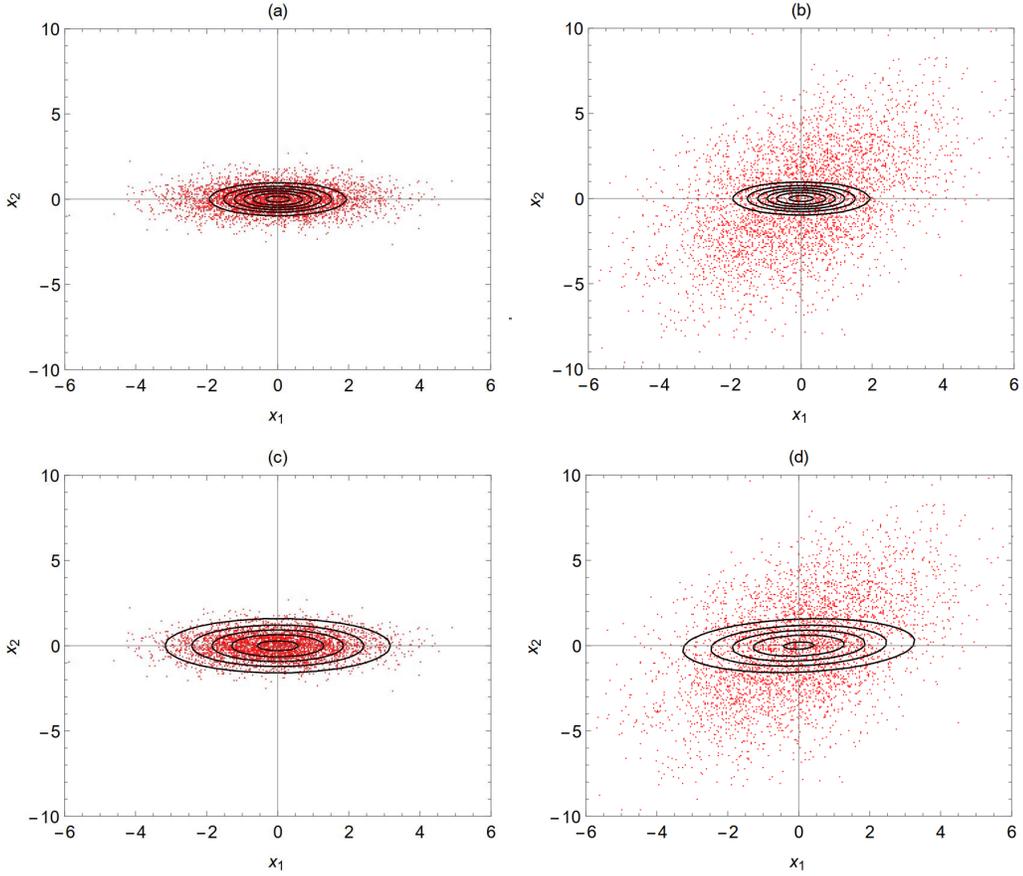

Figure 2: (color online) Set of 4000 points generated from the joint probability distribution $\rho(x_1, x_2, \tau) = \int_{-\infty}^{+\infty} |\Psi(x_1, x_2, x_3, \tau)|^2 \, dx_3$ with the wave function, Eq. (2.11), at time of measurement $t = \tau = 1/\kappa$ using the values $\delta_q = 1$, $b = 2\delta_q^2$, and $m_i = 1$. Each point constitutes a joint record of the pointers, which is registered in the $x_1 x_2$-plane. To graphically inspect the notion of retrodictive and predictive accuracy, the whole set of records is compared versus the contour plots of the Wigner function (with $\delta_q = 1$) associated to the Gaussian state under inspection before (Figs. (a), (b)) and after (Figs. (c), (d)) the measurement respectively, for the cases of strong coupling ($\kappa = 100$, left Figs.) and weak coupling ($\kappa = 0.5$, right Figs.). The deviation from the Wigner functions becomes larger in the weak coupling regime. Notably, the correlation (the rotation angle of the statistical distribution of the joint readings) between the pointer records is greater as the coupling strength is weaker.

The definition of the predictive aspect of accuracy is quite important in the quantum description of a measurement process, unlike classical mechanics where these effects can be taken depreciable. Concerning the Arthurs-Kelly model, the consideration of the retrodictive and predictive aspects of accuracy allow to quantitatively define the concept of a disturbance operator [66] like the one defined in [63], which, through an adequate measure of dispersion, enable to quantify the feedback on the conjugate probability distributions of the system due to the measurement process [8].

Hence the two aspects of accuracy in the simultaneous measurement will be decreasing as the coupling strength, $\kappa$, becomes smaller; this fact will be reflected in the deviation of the statistical distributions of the joint readings from those of the canonical pair before and after the measurement.

To graphically illustrate that argument, in Fig. 2 we plot the set of points coming from the joint probability density

$$\rho(x_1, x_2, \tau) = \int_{-\infty}^{+\infty} |\Psi(x_1, x_2, x_3, \tau)|^2\, dx_3, \qquad (3.20)$$

which represent the distribution of the pointer readings at the time of the measurement; then, we compare it versus both, the Wigner quasi-probability distribution of the Gaussian system before (retrodictive aspect) the measurement

$$W(x_3, p_3, t=0) = \pi^{-1} \int_{-\infty}^{+\infty} \phi_3^*(x_3+\xi)\phi_3(x_3-\xi)e^{2ip_3\xi}d\xi, \qquad (3.21)$$

and just after (predictive aspect) the measurement

$$W(x_3, p_3, t=\tau) = \pi^{-1} \int_{-\infty}^{+\infty} \Psi^*(x_1,x_2,x_3+\xi,\tau)\Psi(x_1,x_2,x_3-\xi,\tau)e^{2ip_3\xi}\, dx_1 dx_2 d\xi; \qquad (3.22)$$

therefore, the maximal degree of match is determined within the margins stipulated by the variances of the error operators, Eqs. (3.16) to (3.19).

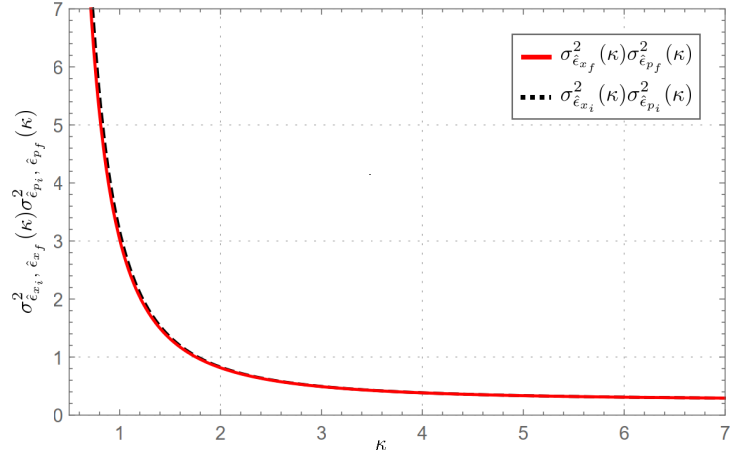

Figure 3: (color online) Behavior of the products $\sigma^2_{\hat{\epsilon}_{x_i}}(\kappa)\,\sigma^2_{\hat{\epsilon}_{p_i}}(\kappa)$ (dashed plot) and $\sigma^2_{\hat{\epsilon}_{x_f}}(\kappa)\,\sigma^2_{\hat{\epsilon}_{p_f}}(\kappa)$ (continuous plot) given by Eqs. (3.26) and (3.27) versus the coupling constant $\kappa$ for the values $\delta_q = 1$, $m_i = 1$. As the coupling becomes smaller, the free evolution of the measurement setting (quantified by the $\Delta_{1,2}(\kappa)$-functions) causes an hyperbolic shift upwards of the products of variances of the retrodictive and predictive error operator from its minimal values.

Hence, the minimum product of pointer variances found by Arthurs and Kelly is recovered at the strong coupling regime, that is,

$$\lim_{\kappa \longrightarrow \infty} \Delta_1(\kappa) = 0. \qquad (3.23)$$

On the other hand, the first one recognizing a lower bounding similar to the one of the Kennard uncertainty relation for the product of variances of the retrodictive errors operators was Arthurs and Kelly; they show that [10]

$$\sigma^2_{\hat{\epsilon}_{x_i}}\sigma^2_{\hat{\epsilon}_{p_i}} \geq \frac{1}{4}. \qquad (3.24)$$

The above inequality is known as the Heisenberg uncertainty relation for joint measurements [63], which is specifically defined for the position and momentum observables. Its proper interpretation goes according to the statute (B) defined in [67], which establishes that although it is impossible to perform a (retrodictive) exact simultaneous measurement of position and momentum is still possible to make an approximate estimation, where the product of the variances of the (retrodictive) error operators will be constrained by Ineq. (3.24). There is a similar interpretation for the inequality bounding the variances of the predictive error operators of the simultaneous measurement, which has shown to be [8,26,27]

$$\sigma^2_{\hat{\epsilon}_{x_f}} \sigma^2_{\hat{\epsilon}_{p_f}} \geq \frac{1}{4}. \tag{3.25}$$

Hence from the product of variances, Eqs. (3.6) and (3.7), it is possible to see that an optimal retrodictive and predictive simultaneous measurement saturates the inequalities Eqs. (3.24) and (3.25) respectively.

By taking the product of Eqs. (3.16) and (3.17), and using the value of the balance parameter $b = 2\delta_q^2$ as well as the lower bound of Kennard uncertainty relation, we obtain the value of the minimum product of variances of the retrodictive error operators plus a $\kappa$-dependent contribution, that is,

$$\sigma^2_{\hat{\epsilon}_{x_i}}(\kappa) \sigma^2_{\hat{\epsilon}_{p_i}}(\kappa) = \left\{\sigma^2_{\hat{\epsilon}_{x_i}} \sigma^2_{\hat{\epsilon}_{p_i}}\right\}_{\min} + \Delta_2(\kappa). \tag{3.26}$$

In the same way, by taking the product of Eqs. (3.18) and (3.19) and following the same procedure above, we obtain a similar result for the product of variances of the predictive error operators

$$\sigma^2_{\hat{\epsilon}_{x_f}}(\kappa) \sigma^2_{\hat{\epsilon}_{p_f}}(\kappa) = \left\{\sigma^2_{\hat{\epsilon}_{x_f}} \sigma^2_{\hat{\epsilon}_{p_f}}\right\}_{\min} + \Delta_3(\kappa), \tag{3.27}$$

where $\left\{\sigma^2_{\hat{x}_i} \sigma^2_{\hat{p}_i}\right\}_{\min} = 1/4 = \left\{\sigma^2_{\hat{x}_f} \sigma^2_{\hat{p}_f}\right\}_{\min}$ and the $\Delta_{1,2}(\kappa)$-functions are defined as

$$\Delta_2(\kappa) = \frac{11m_1^2\left(8\delta_q^4 + \kappa^2 m_2^2\right) - 24m_1\left(8\delta_q^4 + \kappa^2 m_2^2\right)m_3 + 72\left(8\delta_q^4 + \kappa^2\left[16\delta_q^8 m_1^2 + m_2^2\right]\right)m_3^2}{\left(24\delta_q^2 \kappa^2 m_1 m_2 m_3\right)^2}, \tag{3.28}$$

$$\Delta_3(\kappa) = \frac{m_1^2\left(8\delta_q^4 + \kappa^2 m_2^2\right) - 12m_1\left(8\delta_q^4 + \kappa^2 m_2^2\right)m_3 + 36\left(8\delta_q^4 + \kappa^2\left[16\delta_q^8 m_1^2 + m_2^2\right]\right)m_3^2}{288\left(\delta_q^2 \kappa^2 m_1 m_2 m_3\right)^2}, \tag{3.29}$$

which are zero in the ideal situation of strong coupling, therefore

$$\lim_{\kappa \longrightarrow \infty} \Delta_{2,3}(\kappa) = 0, \tag{3.30}$$

recovering the lower bounds of the retrodictive and predictive uncertainty relations, Eqs. (3.24) and (3.25). Therefore, the products $\sigma^2_{\hat{\epsilon}_{x_i}}(\kappa) \sigma^2_{\hat{\epsilon}_{p_i}}(\kappa)$ and $\sigma^2_{\hat{\epsilon}_{x_f}}(\kappa) \sigma^2_{\hat{\epsilon}_{p_f}}(\kappa)$ are hyperbolically shifted from its minimum values as the coupling strength becomes smaller; see Fig. 3. Therefore the retrodictive and predictive experimental errors are higher as the coupling strength $\kappa$ is small.

## 4 Conclusions

In this paper, we investigate the effects caused by the free dynamics in the Arthurs-Kelly's model of simultaneous measurement and its consequences on the retrodictive and predictive aspects of the accuracy of the measurement process. One of our main results is that the inclusion of the free energy Hamiltonian generates a spreading of the whole probability density, resulting in an extra noise compared to the one that is induced by the measurement process in the optimal regime of accuracy. That is to say, the free evolution increases the statistical distributions of the pointer readings; this results in a $\kappa$-dependent increment of the variances of the retrodictive and predictive error operators.

For clarity, we exemplify these arguments through the analysis of two limit situations. (i) Strong coupling regime. Where, due to the reciprocal relationship between the time of the measurement process and the coupling strength $\kappa$, the measurement can be taken as instantaneous; consequently, the spread affecting the whole probability density is depreciable. Hence, the variances of the retrodictive and predictive error operators are minimum; then, the optimal accuracy in both aspects can be reached. (ii) Weak coupling regime. Where the pointers are weakly coupled with the Gaussian system to be inspected; hence, the measurement takes a considerable time-lapse and the free energy operators must be taken into account in the Hamiltonian governing the dynamics; thus, the spread of the whole probability density becomes substantial, deviating the product of the variances of the retrodictive and predictive error operators far from their minimal values; thus, the two aspects of the accuracy are small.

In conclusion, taking into account the full dynamics, in the measurement process of two non-conmuting observables, the (optimal) retrodictive and predictive accuracy diminish as the coupling strength between the measuring apparatuses and the gaussian system become smaller. However, we establish the fact that the strength of this conclusion should be maintained for any other state over which the measurement was carried out, because the free energy operators $e^{a(t)\hat{P}^2}$ appearing in the time evolution operator, Eq. (2.10), will cause its free-propagation[2], i. e., the temporal dispersion of its associated probability distribution [68, 69].

# Appendices

## A  Factorization of The Time Evolution Operator

In this appendix, we show the process for expressing the time evolution operator, Eq. (2.10), as a product of exponential operators; hence we follow the methodology exposed in [45]. With the Hamiltonian, Eq. (2.8), the time evolution operator is

$$e^{-i\hat{H}t} = e^{-\frac{it}{2m_1}\hat{p}_1^2} e^{-\frac{it}{2m_2}\hat{p}_2^2} e^{-it\left(\frac{\hat{p}_3^2}{2m_3} + \kappa\hat{x}_3\hat{p}_1 + \kappa\hat{p}_3\hat{p}_2\right)}, \tag{A.1}$$

hence it is necessary to factor the last exponential in the above equation because $\hat{x}_3$ do not commute with $\hat{p}_3$ and $\hat{p}_3^2$. Take the operators

$$\hat{A} = -it\left(\frac{\hat{p}_3^2}{2m_3} + \kappa\hat{p}_3\hat{p}_2\right), \tag{A.2}$$

$$\hat{B} = -it\kappa\hat{x}_3\hat{p}_1; \tag{A.3}$$

then, the following commutators are obtained through a test function $f(x_3)$

$$\left[\hat{A}, \hat{B}\right] = it^2\left(\frac{\kappa}{m_3}\hat{p}_1\hat{p}_3 + \kappa^2\hat{p}_1\hat{p}_2\right), \tag{A.4}$$

$$\left[\hat{A}, \left[\hat{A}, \hat{B}\right]\right] = 0, \tag{A.5}$$

$$\left[\hat{B}, \left[\hat{A}, \hat{B}\right]\right] = \frac{it^3\kappa^2}{m_3}\hat{p}_1^2. \tag{A.6}$$

We now define an auxiliary function $F(\zeta)$ in terms of the exponential to factor

$$F(\zeta) = e^{\zeta(\hat{A}+\hat{B})}, \tag{A.7}$$

and its derivative

$$F'(\zeta) = (\hat{A} + \hat{B})F(\zeta), \tag{A.8}$$

where $\zeta$ is an auxiliary parameter which we take as 1 after finish all calculations; we then set a generic factorization that allows factorizing the last exponential of Eq. (A.1) in the most convenient manner; that is

$$F(\zeta) = e^{f_0(\zeta)} e^{f_1(\zeta)\hat{N}_1} e^{f_2(\zeta)\hat{N}_2} e^{f_3(\zeta)\hat{N}_3}; \tag{A.9}$$

---
[2]it can be proved through the convolution theorem for inverse Fourier transform, that the application of this operator to an arbitrary state is equivalent to the dynamics with the free particle-propagator in the method of Green's functions.

besides, its derivative is

$$F'(\zeta) = f_0'(\zeta)F(\zeta) + f_1'(\zeta)\hat{N}_1 F(\zeta) + f_2'(\zeta)e^{f_0(\zeta)}e^{f_1(\zeta)\hat{N}_1}\hat{N}_2 e^{f_2(\zeta)\hat{N}_2}e^{f_3(\zeta)\hat{N}_3}$$
$$+ f_3'(\zeta)e^{f_0(\zeta)}e^{f_1(\zeta)\hat{N}_1}e^{f_2(\zeta)\hat{N}_2}\hat{N}_3 e^{f_3(\zeta)\hat{N}_3}. \quad (A.10)$$

In this work we consider $\hat{N}_1 = \hat{A}$, $\hat{N}_2 = \hat{B}$, $\hat{N}_3 = \hat{A}$; therefore, Eq. (A.10) is

$$F'(\zeta) = f_0'(\zeta)F(\zeta) + f_1'(\zeta)\hat{A}F(\zeta) + f_2'(\zeta)e^{f_0(\zeta)}e^{f_1(\zeta)\hat{A}}e^{f_2(\zeta)\hat{B}}e^{f_3(\zeta)\hat{A}}$$
$$+ f_3'(\zeta)e^{f_0(\zeta)}e^{f_1(\zeta)\hat{A}}e^{f_2(\zeta)\hat{B}}\hat{A}e^{f_3(\zeta)\hat{A}}; \quad (A.11)$$

then, according to Eq. (A.8), it is necessary to factorize the function $F(\zeta)$ to the right in the right-hand side of the above equation, but the third and fourth terms prevent it. The third term of Eq. (A.11) is solved as

$$f_2'(\zeta)e^{f_0(\zeta)}e^{f_1(\zeta)\hat{A}}\hat{B}e^{f_2(\zeta)\hat{B}}e^{f_3(\zeta)\hat{A}} = f_2'(\zeta)\left(\hat{B} + f_1(\zeta)\left[\hat{A},\hat{B}\right]\right)F(\zeta), \quad (A.12)$$

where we have used the fact that for self-adjoint operators $\hat{X}$ and $\hat{Y}$ it is satisfied that $e^{f_i(\zeta)\hat{X}}\hat{Y} = \left(e^{f_i(\zeta)\hat{X}}\hat{Y}e^{-f_i(\zeta)\hat{X}}\right)e^{f_i(\zeta)\hat{X}}$, together the commutators, Eqs. (A.4) and (A.5), and the condition [70]:

$$\left(e^{f_i(\zeta)\hat{X}}\hat{Y}e^{-f_i(\zeta)\hat{X}}\right) = \hat{Y} + f_i\left[\hat{X},\hat{Y}\right] + \frac{f_i^2}{2!}\left[\hat{X},\left[\hat{X},\hat{Y}\right]\right] + \cdots, \quad (A.13)$$

following the same procedure and using the commutators Eqs. (A.4) and (A.6), the fourth therm of Eq. (A.11) is given by

$$f_3'(\zeta)e^{f_0(\zeta)}e^{f_1(\zeta)\hat{A}}e^{f_2(\zeta)\hat{B}}\hat{A}e^{f_3(\zeta)\hat{A}} = f_3'(\zeta)\left(\hat{A} - f_2(\zeta)\left[\hat{A},\hat{B}\right] - \frac{(f_2(\zeta))^2}{2}\left[\hat{B},\left[\hat{A},\hat{B}\right]\right]\right)F(\zeta); \quad (A.14)$$

therefore the derivative, Eq. (A.11), is expressed as

$$F'(\zeta) = \left\{f_0'(\zeta) + f_1'(\zeta)\hat{A} + f_2'(\zeta)\left(\hat{B} + f_1(\zeta)[\hat{A},\hat{B}]\right)\right.$$
$$\left. + f_3'(\zeta)\left(\hat{A} - f_2(\zeta)[\hat{A},\hat{B}] - (f_2^2(\zeta)/2)[\hat{B},[\hat{A},\hat{B}]]\right)\right\}F(\zeta), \quad (A.15)$$

equalling it with Eq. (A.8), we obtain the following set of differential equations

$$f_1'(\zeta) + f_3'(\zeta) = 1, \quad (A.16)$$

$$f_2'(\zeta) = 1, \quad (A.17)$$

$$f_0'(\zeta) - \frac{f_3'(\zeta)f_2^2(\zeta)}{2}[\hat{B},[\hat{A},\hat{B}]] = 0, \quad (A.18)$$

$$f_2'(\zeta)f_1(\zeta) - f_3'(\zeta)f_2(\zeta) = 0, \quad (A.19)$$

which are subject to the conditions

$$f_0(0) = f_1(0) = f_2(0) = f_3(0) = 0, \quad (A.20)$$

the solutions are given by

$$f_0(\zeta) = \frac{[\hat{B},[\hat{A},\hat{B}]]}{12}\zeta^3, \quad (A.21)$$

$$f_1(\zeta) = \frac{\zeta}{2}, \quad (A.22)$$

$$f_2(\zeta) = \zeta, \quad (A.23)$$

$$f_3(\zeta) = \frac{\zeta}{2}. \quad (A.24)$$

Substituting Eqs. (A.21) to (A.24) in Eq. (A.9) with $\hat{N}_1 = \hat{A}$, $\hat{N}_2 = \hat{B}$, $\hat{N}_3 = \hat{A}$ and doing $\zeta = 1$, the time evolution operator, Eq. (A.1), is expressed as

$$e^{-i\hat{H}t} = e^{\Delta x_1 \hat{p}_1^2} e^{-\frac{it}{2m_2}\hat{p}_2^2} e^{-\frac{it}{4m_3}\hat{p}_3^2} e^{-\frac{it\kappa}{2}\hat{p}_3\hat{p}_2} e^{-it\kappa\hat{x}_3\hat{p}_1} e^{-\frac{it}{4m_3}\hat{p}_3^2} e^{-\frac{it\kappa}{2}\hat{p}_3\hat{p}_2}, \tag{A.25}$$

where we have grouped the operators with power $\hat{p}_1^2$ into one, and we have taken

$$\Delta x_1 = -(it/2m_1) + (it^3\kappa^2/12m_3). \tag{A.26}$$

# B Time-dependent coefficients of the wave function describing the dynamics of the simultaneous measurement

In this appendix, we define the form of the time-dependent functions $\mathcal{N}(t)$ and $\varepsilon_j(t)$ appearing in the wave function, Eq. (2.11), which dictates the temporal evolution through the measurement process of the whole configuration.

The $\varepsilon_j(t)$ functions are defined as

$$\varepsilon_j(t) = \frac{\Gamma_j(t)}{\Theta(t)}, \quad j = 1, 2, 3, 4, 5, 6, \tag{B.1}$$

with

$$\Gamma_1(t) = 3m_1m_3\left(t\left(-2bt + m_2\left(i + 4b\kappa^2 m_3 t\right)\right) + 4m_3\left(m_2 + 2ibt\right)\delta_q^2\right), \tag{B.2}$$

$$\Gamma_2(t) = bm_2\left(3bm_1m_3 + i\left(-6m_3t + \kappa^2 m_1 t^3\right)\right)\left(it + 4m_3\delta_q^2\right), \tag{B.3}$$

$$\Gamma_3(t) = \quad m_3\left(6ib^2m_1m_3t + 2m_2t\left(2i\kappa^2 m_1 t^2 + m_3\left(-3i + 6\kappa^2 m_1 t \delta_q^2\right)\right)\right.$$
$$\left. + b\left(12m_3t^2 + m_1\left(3m_2\left(m_3 + \kappa^4 m_3 t^4\right) - 8\kappa^2 t^3\left(t - 3im_3\delta_q^2\right)\right)\right)\right), \tag{B.4}$$

$$\Gamma_4(t) = 6b\kappa^2 m_1 m_2 m_3 t^2\left(-t + 4im_3\delta_q^2\right), \tag{B.5}$$

$$\Gamma_5(t) = -6\kappa m_1 m_3 t\left(t\left(im_2 + 2b\left(-1 + \kappa^2 m_2 m_3\right)t\right) + 4m_3\left(m_2 + 2ibt\right)\delta_q^2\right), \tag{B.6}$$

$$\Gamma_6(t) = 2b\kappa m_2 m_3 t\left(-6ibm_1m_3 + t\left(5k^2 m_1 t^2 + m_3\left(-12 - 12i\kappa^2 m_1 t \delta_q^2\right)\right)\right); \tag{B.7}$$

and

$$\Theta(t) = ibt\left(3m_1m_2m_3 + 12m_3\left(1 - 2\kappa^2 m_2 m_3\right)t^2 + \kappa^2 m_1\left(-2 + 7\kappa^2 m_2 m_3\right)t^4\right)$$
$$+ 4bm_3\left(3m_1m_2m_3 + 12m_3t^2 + \kappa^2 m_1\left(-2 + 3\kappa^2 m_2 m_3\right)t^4\right)\delta_q^2$$
$$+ m_2t\left(6m_3 - \kappa^2 m_1 t^2\right)\left(t - 4im_3\delta_q^2\right) + 6b^2 m_1 m_3 t\left(\left(-1 + 2\kappa^2 m_2 m_3\right)t + 4im_3\delta_q^2\right). \tag{B.8}$$

While the $\mathcal{N}(t)$-function is given by

$$\mathcal{N}(t) = 2\left(\frac{2b^2\delta_q^2}{\pi^3}\right)^{\frac{1}{4}}\left(\frac{3m_1m_2m_3^2}{\Theta(t)}\right)^{\frac{1}{2}}, \tag{B.9}$$

which plays the role of a normalization constant.

# C  Application of the time evolution operator

Because the initial function, Eq. (2.5), is in the canonical position representation, it is convenient to express the time evolution operator also in the same basis, this is done by simply taking the substitution $\hat{p}_{i'} \to -i\partial_{x_{i'}}$ for $i' = 1, 2, 3$. Thus, the factorization given by Eq. (A.25), can be expressed as

$$e^{-i\hat{H}t} = e^{-\Delta x_1 \partial_{x_1}^2} e^{\delta_2(t)\partial_{x_2}^2} e^{\frac{\delta_3(t)}{2}\partial_{x_3}^2} e^{\delta_\kappa(t)\partial_{x_3}\partial_{x_2}} e^{\delta'_k(t)\hat{x}_3\partial_{x_1}} e^{\frac{\delta_3(t)}{2}\partial_{x_3}^2} e^{\delta_\kappa(t)\partial_{x_3}\partial_{x_2}}, \tag{C.1}$$

where we have done

$$\delta_{i'}(t) = \frac{it}{2m_{i'}}, \qquad i' = 1, 2, 3. \tag{C.2}$$

And

$$\delta_\kappa(t) = it\kappa/2, \tag{C.3}$$

$$\delta'_\kappa(t) = -t\kappa. \tag{C.4}$$

We will refer to the exponentials of Eq. (C.1) from right to left, labeling them from the first to seventh respectively; to apply them, we use the one and the two dimensional Fourier transform (FT)

$$\mathcal{F}[f(q)](p) = g(p) = \frac{1}{\sqrt{2\pi}} \int_{-\infty}^{+\infty} f(q)e^{-iqp}dq, \tag{C.5}$$

$$\mathcal{F}_{2\mathbb{D}}[m(x,y)](u,v) = n(u,v) = \frac{1}{2\pi} \int_{-\infty}^{+\infty} \int_{-\infty}^{+\infty} m(x,y)e^{-i(xu+yv)}dxdy, \tag{C.6}$$

and the one and two dimensional inverse Fourier transform (IFT)

$$\mathcal{F}^{-1}[g(p)](q) = f(q) = \frac{1}{\sqrt{2\pi}} \int_{-\infty}^{+\infty} g(p)e^{iqp}dp, \tag{C.7}$$

$$\mathcal{F}_{2\mathbb{D}}^{-1}[n(u,v)](x,y) = m(x,y) = \frac{1}{2\pi} \int_{-\infty}^{+\infty} \int_{-\infty}^{+\infty} n(u,v)e^{i(xu+yv)}dudv; \tag{C.8}$$

besides, we use the expansion in McLaurin series of an exponential operator when we apply it to a wave function, that is,

$$e^{\hat{A}}\Psi = \sum_{n=0}^{\infty} \frac{1}{n!} \hat{A}^n \Psi. \tag{C.9}$$

Applying the first operator of Eq. (C.1) to the initial function, Eq. (2.5), we have

$$\psi_1 = e^{\delta_\kappa(t)\partial_{x_3}\partial_{x_2}} \phi_1(x_1)\phi_2(x_2)\phi_3(x_3) = \phi_1(x_1) \sum_{n=0}^{\infty} \frac{(\delta_\kappa(t))^n}{n!} \frac{d^n \phi_3(x_3)}{dx_3^n} \frac{d^n \phi_2(x_2)}{dx_2^n}, \tag{C.10}$$

Applying two dimensional FT in $x_2$ and $x_3$ variables to Eq. (C.10), we have

$$\mathcal{F}_{2\mathbb{D}}[\psi_1](x_1, p_2, p_3, t)$$

$$= \phi_1(x_1) \sum_{n=0}^{\infty} \frac{(\delta_\kappa(t))^n}{n!} \mathcal{F}\left[\frac{d^n \phi_2(x_2)}{dx_2^n}\right](p_2) \mathcal{F}\left[\frac{d^n \phi_3(x_3)}{dx_3^n}\right](p_3)$$

$$= \phi_1(x_1) \sum_{n=0}^{\infty} \frac{(\delta_\kappa(t))^n}{n!} (ip_2)^n (ip_3)^n \mathcal{F}[\phi_2(x_2)](p_2) \mathcal{F}[\phi_3(x_3)](p_3)$$

$$= \phi_1(x_1) e^{-\delta_\kappa(t)p_2 p_3} \mathcal{F}[\phi_2(x_2)](p_2) \mathcal{F}[\phi_3(x_3)](p_3), \tag{C.11}$$

where in the second line we have used the case of separable functions (see for example [71] pp. 9-10), and in the third line the derivative theorem of the one FT

$$\mathcal{F}\left[\frac{d^n f(q)}{dq^n}\right](p) = (ip)^n \mathcal{F}[f(q)](p). \tag{C.12}$$

Using the definition, Eq. (2.4), for $\phi_3(x_3)$ and $\phi_2(x_2)$ and taking the two-dimensional IFT, the application of the first operator is finished, the result is

$$\psi_1 = \left(\frac{\delta_q}{(2\pi)^{1/2}\sigma_1(t)}\right)^{\frac{1}{2}} \phi_1(x_1)\phi_2(x_2)\, e^{-\frac{(x_3 - 2b\delta_\kappa(t)x_2)^2}{4\sigma_1(t)}}, \tag{C.13}$$

with

$$\sigma_1(t) = \left(\delta_q^2 - b\left(\delta_\kappa(t)\right)^2\right). \tag{C.14}$$

As can be seen, the wave function has become entangled between the position variables $x_2$ and $x_3$; therefore, the initial product structure has been lost due to the unitary dynamics.

Following a similar methodology, we apply the second operator of Eq. (C.1) Eq. (C.13)

$$\psi_2 = e^{\frac{\delta_3(t)}{2}\partial_{x_3}^2}\psi_1 = \psi_1' \sum_{n=0}^{\infty} \frac{\left(\frac{\delta_3(t)}{2}\right)^n}{n!} \partial_{x_3}^{2n} \left\{ (\sigma_1(t))^{-1/2} e^{-\frac{(x_3 - 2b\delta_\kappa(t)x_2)^2}{4\sigma_1(t)}} \right\}, \tag{C.15}$$

where we have left inside the brackets the term $(\sigma_1(t))^{-1/2}$, in order to simplify the application of the exponential operator (see for example the methodology exposed in [72]) and we take

$$\psi_1' = \left(\frac{\delta_q}{(2\pi)^{1/2}}\right)^{\frac{1}{2}} \phi_1(x_1)\phi_2(x_2). \tag{C.16}$$

Applying FT in $x_3$ variable to Eq. (C.15), we have

$$\mathcal{F}[\psi_2](x_1, x_2, p_3, t)$$

$$= \psi_1' \sum_{n=0}^{\infty} \frac{\left(\frac{\delta_3(t)}{2}\right)^n}{n!} \mathcal{F}\left[\partial_{x_3}^{2n}\left\{(\sigma_1(t))^{-1/2} e^{-\frac{(x_3 - 2b\delta_\kappa(t)x_2)^2}{4\sigma_1(t)}}\right\}\right](x_1, x_2, p_3, t)$$

$$= \psi_1' \sum_{n=0}^{\infty} \frac{\left(\frac{\delta_3(t)}{2}\right)^n}{n!} (ip_3)^{2n} \mathcal{F}\left[\left\{(\sigma_1(t))^{-1/2} e^{-\frac{(x_3 - 2b\delta_\kappa(t)x_2)^2}{4\sigma_1(t)}}\right\}\right](x_1, x_2, p_3, t)$$

$$= \psi_1' e^{-\left(\frac{\delta_3(t)}{2}\right)p_3^2} \mathcal{F}\left[\left\{(\sigma_1(t))^{-1/2} e^{-\frac{(x_3 - 2b\delta_\kappa(t)x_2)^2}{4\sigma_1(t)}}\right\}\right](x_1, x_2, p_3, t), \tag{C.17}$$

Taking the IFT in $p_{x_3}$ variable to Eq. (C.17), the application of the second operator is finished, the result is

$$\psi_2 = \left(\frac{\delta_q}{(2\pi)^{1/2}\sigma_2(t)}\right)^{\frac{1}{2}} \phi_1(x_1)\phi_2(x_2)\, e^{\frac{-(x_3 - 2b\delta_\kappa(t)x_2)^2}{4\sigma_2(t)}}, \tag{C.18}$$

with

$$\sigma_2(t) = \left(\delta_q^2 - b\left(\delta_\kappa(t)\right)^2 + \frac{\delta_3(t)}{2}\right); \tag{C.19}$$

then, the effect of the second operator is to increase the amplitude of the entangled wave function and to contribute to the dispersion of the Gaussian wave packet associated with the entangled variables $x_2$ and $x_3$. In general, the operators of the type $e^{C\partial_{x_i}^2}$ have this dispersive effect when they act on Gaussian wave packets; see [72].

We now apply the third operator

$$\psi_3 = e^{\delta_k'(t)\hat{x}_3\partial_{x_1}}\psi_2 = \sum_{n=0}^{\infty} \frac{(\delta_k'(t))^n}{n!} \hat{x}_3^n \partial_{x_1}^n \psi_2 = \psi_2' \sum_{n=0}^{\infty} \frac{(\delta_k'(t))^n}{n!} x_3^n \left\{\partial_{x_1}^n \phi_1(x_1)\right\}, \tag{C.20}$$

where we have done

$$\psi_2' = \left(\frac{\delta_q}{(2\pi)^{1/2}\sigma_2(t)}\right)^{\frac{1}{2}} \phi_2(x_2)\, e^{\frac{-(x_3 - 2b\delta_\kappa(t)x_2)^2}{4\sigma_2(t)}}. \tag{C.21}$$

Taking the FT of Eq. (C.20) in $x_1$ variable

$$\mathcal{F}\left[\psi_3\right](p_1, x_2, x_3, t) \tag{C.22}$$

$$= \psi_2' \sum_{n=0}^{\infty} \frac{(\delta_k'(t)\, x_3)^n}{n!} \mathcal{F}\left[\partial_{x_1}^n \phi_1(x_1)\right](p_1, x_2, x_3, t)$$

$$= \psi_2' \sum_{n=0}^{\infty} \frac{(i\delta_k'(t)\, x_3 p_1)^n}{n!} \mathcal{F}\left[\phi_1(x_1)\right](p_1, x_2, x_3, t)$$

$$= \psi_2' e^{i\delta_k'(t) x_3 p_1} \mathcal{F}\left[\phi_1(x_1)\right](p_1, x_2, x_3, t), \tag{C.23}$$

Taking the IFT on Eq. (C.23) in $p_1$ variable, the application of the third operator is finished; thus we have

$$\psi_3 = \left(\frac{\delta_q}{(2\pi)^{1/2}\sigma_2(t)}\right)^{\frac{1}{2}} \phi_1\left(x_1 + \delta_k'(t)\, x_3\right) \phi_2(x_2)\, e^{\frac{-(x_3 - 2b\delta_\kappa(t) x_2)^2}{4\sigma_2(t)}}. \tag{C.24}$$

The effect of the third operator is proportionally to displace the $x_1$ variable by the $x_3$ variable or equivalently to entangled them; thus, this operator acts as a displacement operator.

It must be noted that all operators in Eq. (C.1) that are still to apply commute, hence the order of application of they does not matter from this point. Following this idea, we apply the seventh operator of Eq. (C.1) to Eq. (C.24)

$$\psi_7 = e^{-\Delta x_1(t)\, \partial_{x_1}^2} \psi_3 = \psi_3' \sum_{n=0}^{\infty} \frac{(-\Delta x_1(t))^n}{n!} \partial_{x_1}^{2n} \phi_1\left(x_1 + \delta_k'(t)\, x_3\right), \tag{C.25}$$

where we have taken

$$\psi_3' = \left(\frac{\delta_q}{(2\pi)^{1/2}\sigma_2(t)}\right)^{\frac{1}{2}} \phi_2(x_2)\, e^{\frac{-(x_3 - 2b\delta_\kappa(t)\, x_2)^2}{4\sigma_2(t)}}. \tag{C.26}$$

Taking the FT of Eq. (C.25) in $x_1$ variable, we have

$$\mathcal{F}\left[\psi_7\right](p_1, x_2, x_3, t)$$

$$= \psi_3' \sum_{n=0}^{\infty} \frac{(-\Delta x_1(t))^n}{n!} \mathcal{F}\left[\partial_{x_1}^{2n} \phi_1(x_1 + \delta_k'(t)\, x_3)\right](p_1, x_2, x_3, t)$$

$$= \psi_3' \sum_{n=0}^{\infty} \frac{(-\Delta x_1(t))^n (ip_1)^{2n}}{n!} \mathcal{F}\left[\phi_1(x_1 + \delta_k'(t)\, x_3)\right](p_1, x_2, x_3, t)$$

$$= \psi_3' e^{\Delta x_1(t)\, p_1^2} \mathcal{F}\left[\phi_1(x_1 + \delta_k'(t)\, x_3)\right](p_1, x_2, x_3, t) \tag{C.27}$$

Taking the definition for $\phi_1\left(x_1 + \delta_k'(t)\, x_3\right)$ (that is, Eq. (2.4) proportionally displaced by $x_3$) and the IFT in $p_1$ variable; we finish the application of the seventh operator, the result is

$$\psi_7 = \left(\frac{2b^2\delta_q^2}{\pi^3}\right)^{\frac{1}{4}} (\sigma_2(t)\sigma_3(t))^{-\frac{1}{2}}\, e^{-bx_2^2} e^{-\frac{\left(x_1 + \delta_k'(t)\, x_3\right)^2}{\sigma_3(t)}} e^{\frac{-(x_3 - 2b\delta_\kappa(t)\, x_2)^2}{4\sigma_2(t)}}, \tag{C.28}$$

with

$$\sigma_3(t) = b + 4\Delta x_1(t). \tag{C.29}$$

The seventh operator contributes to the amplitude and the dispersion of the wave function associated with $x_1$ and $x_3$ variables. Before applying the next operator, we complete the square binomial in the power of exponentials for the $x_2$ variable; thus, the wave function, Eq. (C.28), can be rewritten as

$$\psi_7 = \left(\frac{2b^2\sigma^2}{\pi^3}\right)^{\frac{1}{4}} (\sigma_2(t)\sigma_3(t))^{-\frac{1}{2}}\, e^{-\frac{(\beta(t)x_2 - \alpha(t)\, x_3)^2}{4\sigma_2(t)}} e^{-\frac{\left(x_1 + \delta_k'(t)x_3\right)^2}{\sigma_3(t)}} e^{-\frac{x_3^2\left(1 - (\alpha(t))^2\right)}{4\sigma_2(t)}}, \tag{C.30}$$

where we have done

$$\beta(t) = \left[(2b\delta_\kappa(t))^2 + 4\sigma_2(t)b\right]^{\frac{1}{2}}, \tag{C.31}$$

$$\alpha(t) = \frac{2b\delta_\kappa(t)}{\beta(t)}. \tag{C.32}$$

We now apply the sixth operator of Eq. (C.1) to Eq. (C.30)

$$\psi_6 = e^{\delta_2(t)\partial_{x_2}^2}\psi_7 = \psi_7' \sum_{n=0}^{\infty} \frac{(\delta_2(t))^n}{n!} \partial_{x_2}^{2n} \left\{ (\sigma_2(t))^{-\frac{1}{2}} e^{-\frac{(\beta(t)x_2 - \alpha(t)\,x_3)^2}{4\sigma_2(t)}} \right\}, \tag{C.33}$$

with

$$\psi_7' = \left(\frac{2b^2\delta_q^2}{\pi^3}\right)^{\frac{1}{4}} (\sigma_3(t))^{-\frac{1}{2}} e^{-\frac{\left(x_1 + \delta_k'(t)\,x_3\right)^2}{\sigma_3(t)}} e^{-\frac{x_3^2\left(1-\alpha^2(t)\right)}{4\sigma_2(t)}}, \tag{C.34}$$

taking the FT on Eq. (C.33) in $x_2$ variable, we have

$$\begin{aligned}
\mathcal{F}\left[\psi_6\right](x_1, p_2, x_3, t) &= \psi_7' \sum_{n=0}^{\infty} \frac{(\delta_2(t))^n}{n!} \mathcal{F}\left[\partial_{x_2}^{2n} \left\{ (\sigma_2(t))^{-\frac{1}{2}} e^{-\frac{(\beta(t)x_2 - \alpha(t)x_3)^2}{4\sigma_2(t)}} \right\}\right](x_1, p_2, x_3, t) \\
&= \psi_7' \sum_{n=0}^{\infty} \frac{(\delta_2(t))^n}{n!} (ip_2)^{2n} \mathcal{F}\left[\left\{ (\sigma_2(t))^{-\frac{1}{2}} e^{-\frac{(\beta(t)x_2 - \alpha(t)x_3)^2}{4\sigma_2(t)}} \right\}\right](x_1, p_2, x_3, t) \\
&= \psi_7' e^{-\delta_2(t)p_2^2} \mathcal{F}\left[\left\{ (\sigma_2(t))^{-\frac{1}{2}} e^{-\frac{(\beta(t)x_2 - \alpha(t)x_3)^2}{4\sigma_2(t)}} \right\}\right](x_1, p_2, x_3, t),
\end{aligned} \tag{C.35}$$

Taking the IFT on Eq. (C.35) in $p_2$ variable, we finish the application of the sixth operator; thus we have

$$\psi_6 = \left(\frac{2b^2\delta_q^2}{\pi^3}\right)^{\frac{1}{4}} (\sigma_3(t)\sigma_4(t))^{-\frac{1}{2}} e^{-\frac{(\beta(t)x_2 - \alpha(t)x_3)^2}{4\sigma_4(t)}} e^{-\frac{\left(x_1 + \delta_k'(t)x_3\right)^2}{\sigma_3(t)}} e^{-\frac{x_3^2\left(1-\alpha^2(t)\right)}{4\sigma_2(t)}}, \tag{C.36}$$

with

$$\sigma_4(t) = \sigma_2(t) + (\beta(t))^2 \delta_2(t). \tag{C.37}$$

The sixth operator also contributes to the amplitude of the whole wave function and the dispersion of the Gaussian wave packet associate to the $x_2$ and $x_3$ variables. Before applying the next operator, we complete the square binomial in $x_3$ variable in the power of exponentials of Eq. (C.36); thus, the wave function, Eq. (C.36), is rewrite as

$$\psi_6 = \left(\frac{2b^2\delta_q^2}{\pi^3}\right)^{\frac{1}{4}} (\sigma_3(t)\sigma_4(t))^{-\frac{1}{2}} e^{-\left(\lambda(t)x_3 + \frac{\xi_1(x_1, x_2, t)}{2\lambda(t)}\right)^2} e^{\left(\frac{\xi_1(x_1, x_2, t)}{2\lambda(t)}\right)^2} e^{-\xi_2(x_1, x_2, t)}, \tag{C.38}$$

with

$$\lambda(t) = \left(\frac{1-\alpha^2(t)}{4\sigma_2(t)} + \frac{(\delta_k'(t))^2}{\sigma_3(t)} + \frac{(\alpha(t))^2}{4\sigma_4(t)}\right)^{\frac{1}{2}}, \tag{C.39}$$

$$\xi_1(x_1, x_2, t) = \left(\frac{2\delta_k'(t)}{\sigma_3(t)} x_1 - \frac{\alpha(t)\beta(t)}{2\sigma_4(t)} x_2\right), \tag{C.40}$$

$$\xi_2(x_1, x_2, t) = \left(\frac{1}{\sigma_3(t)} x_1^2 + \frac{(\beta(t))^2}{4\sigma_4(t)} x_2^2\right). \tag{C.41}$$

Hence we apply the fifth operator of Eq. (C.1) to Eq. (C.38)

$$\psi_5 = \psi_6' \sum_{n=0}^{\infty} \frac{(\delta_3(t)/2)^n}{n!} \partial_{x_3}^{2n} e^{-\left(\lambda(t)x_3 + \frac{\xi_1(x_1, x_2, t)}{2\lambda(t)}\right)^2}, \tag{C.42}$$

where we have taken

$$\psi_6' = \left(\frac{2b^2\delta_q^2}{\pi^3}\right)^{\frac{1}{4}} (\sigma_3(t)\sigma_4(t))^{-\frac{1}{2}} e^{\left(\frac{\xi_1(x_1,x_2,t)}{2\lambda(t)}\right)^2} e^{-\xi_2(x_1,x_2,t)}, \quad (C.43)$$

taking the FT, Eq. (C.5), on Eq. (C.42) in $x_3$ variable, we have

$$\begin{aligned}
\mathcal{F}[\psi_5](x_1, x_2, p_{x_3}, t) &= \psi_5' \sum_{n=0}^{\infty} \frac{(\delta_3(t)/2)^n}{n!} \mathcal{F}\left[\partial_{x_3}^{2n} e^{-\left(\lambda(t)x_3 + \frac{\xi_1(x_1,x_2,t)}{2\lambda(t)}\right)^2}\right](x_1, x_2, p_3, t) \\
&= \psi_5' \sum_{n=0}^{\infty} \frac{(\delta_3(t)/2)^n (ip_3)^{2n}}{n!} \mathcal{F}\left[e^{-\left(\lambda(t)x_3 + \frac{\xi_1(x_1,x_2,t)}{2\lambda(t)}\right)^2}\right](x_1, x_2, p_{x_3}, t) \\
&= \psi_5' e^{-\left(\frac{\delta_3(t)}{2}\right)p_3^2} \mathcal{F}\left[e^{-\left(\lambda(t)x_3 + \frac{\xi_1(x_1,x_2,t)}{2\lambda(t)}\right)^2}\right](x_1, x_2, p_{x_3}, t), \quad (C.44)
\end{aligned}$$

Taking the IFT on Eq. (C.44) in $p_3$ variable, we finish the application of the fifth operator, the result is

$$\psi_5 = \left(\frac{2b^2\delta_q^2}{\pi^3}\right)^{\frac{1}{4}} (\sigma_3(t)\sigma_4(t)\sigma_5(t))^{-\frac{1}{2}} e^{-\frac{\left(\lambda(t)x_3 + \frac{\xi_1(x_1,x_2,t)}{2\lambda(t)}\right)^2}{\sigma_5(t)}} e^{\left(\frac{\xi_1(x_1,x_2,t)}{2\lambda(t)}\right)^2} e^{-\xi_2(x_1,x_2,t)}, \quad (C.45)$$

with

$$\sigma_5(t) = 1 + 2\left(\lambda(t)\right)^2 \delta_3(t); \quad (C.46)$$

then, the fifth operator contributes to the amplitude, and because of the entanglement of all position variables, it also contributes to the temporal dispersion of the whole Gaussian wave function.

Finally, to obtain the final wave function, we apply the fourth operator of Eq. (C.1) to Eq. (C.45)

$$\Psi(x_1, x_2, x_3, t) = e^{\delta_\kappa(t)\partial_{x_3}\partial_{x_2}} \psi_5 = \sum_{n=0}^{\infty} \frac{(\delta_\kappa(t))^n}{n!} \partial_{x_3}^n \partial_{x_2}^n \psi_5, \quad (C.47)$$

Applying the two dimensional FT to Eq. (C.47) in $x_2$ and $x_3$ variables, we have

$$\begin{aligned}
\mathcal{F}_{2\mathbb{D}}\left[e^{\delta_\kappa \partial_{x_3}\partial_{x_2}} \psi_5\right](x_1, p_2, p_3, t) &= \sum_{n=0}^{\infty} \frac{(\delta_\kappa(t))^n}{n!} \mathcal{F}_{2\mathbb{D}}\left[\partial_{x_3}^n \partial_{x_2}^n \psi_5\right](x_1, p_2, p_3, t) \\
&= \sum_{n=0}^{\infty} \frac{(\delta_\kappa(t))^n}{n!} (ip_2)^n (ip_3)^n \mathcal{F}_{2\mathbb{D}}[\psi_5](x_1, p_2, p_3, t) \\
&= e^{-\delta_\kappa(t)p_2 p_3} \mathcal{F}_{2\mathbb{D}}[\psi_5](x_1, p_2, p_3, t), \quad (C.48)
\end{aligned}$$

where in the second line we use the derivative theorem of the two dimensional FT

$$\mathcal{F}_{2\mathbb{D}}\left[\partial_x^n \partial_y^n f(x,y)\right](u,v) = (iu)^n (iv)^n \mathcal{F}_{2\mathbb{D}}[f(x,y)](u,v). \quad (C.49)$$

Taking the two dimensional IFT of Eq. (C.48) in $p_2$ and $p_3$ variables, we finish the application of the fourth operator, we present the final result as

$$\Psi(x_1, x_2, x_3, t) = \\
\mathcal{N}(t) \exp\left[-\left\{\varepsilon_1(t)x_1^2 + \varepsilon_2(t)x_2^2 + \varepsilon_3(t)x_3^2 + \varepsilon_4(t)x_1 x_2 + \varepsilon_5(t)x_1 x_3 + \varepsilon_6(t)x_2 x_3\right\}\right]. \quad (C.50)$$

The wave function, Eq. (C.50), is clearly a Gaussian wave function with temporal dispersion, that is entangled between the three position variables $x_1$, $x_2$ and $x_3$.